# Studying minijets via the $p_T$ dependence of two-particle correlation in azimuthal angle $\phi$

Xin-Nian Wang

*Nuclear Science Division, Mailstop 70A-3307, Lawrence Berkeley Laboratory*

*University of California, Berkeley, California 94720*[†]

*and*

*Physics Department, Duke University, Durham, NC 27706.*

Following my previous proposal that two-particle correlation functions can be used to resolve the minijet contribution to particle production in minimum biased events of high energy hadronic interactions, I study the $p_T$ and energy dependence of the correlation. Using HIJING Monte Carlo model, it is found that the correlation $c(\phi_1,\phi_2)$ in azimuthal angle $\phi$ between two particles with $p_T > p_T^{\rm cut}$ resembles much like two back-to-back jets as $p_T^{\rm cut}$ increases at high colliding energies due to minijet production. It is shown that $c(0,0) - c(0,\pi)$, which is related to the relative fraction of particles from minijets, increases with energy. The background of the correlation for fixed $p_T^{\rm cut}$ also grows with energy due to the increase of multiple minijet production. Application of this analysis to the study of jet quenching in ultrarelativistic heavy ion collisions is also discussed.

Typeset Using *REVTEX*

# I. INTRODUCTION

Minijets with $p_T \gtrsim$ few GeV/c are commonly believed to become increasingly important in hadronic interactions for energies beyond the CERN Intersecting Storage Ring (ISR) energy range. It has been suggested by many authors that they are responsible for not only the global properties, like the rapid increase of the total cross sections[1]-[7] and the average charged multiplicity[7,10], but also the local correlation and fluctuations[6]-[10] of multiparticle production in high energy $pp$ and $p\bar{p}$ collisions. They have also been estimated to play an important role in ultrarelativistic heavy ion collisions[11].

Another particularly interesting feature of minijets is that they seem to give a natural explaination[12] to the apparent "flow" effect in high energy $p\bar{p}$ collisions. This "flow" effect is the observation in experiments[13] at Fermilab Tevatron Collider energy that the average transverse momentum of charged particles increases with the total multiplicity of the events and the increase is stronger for heavy particles than the light ones. However, this observation is surprisingly in coincident with the results of an equilibrated quark gluon plasma (QGP)[14] in which the common transverse flow velocity from the collective expansion gives heavy particles larger transverse momentum than the pions. The speculation was made even more plausible by Lévai and Müller's finding[15] that there is no time for the baryons to equilibrate with the pions during the expansion of a hadronic fireball. Other models such as the string fusion model[16] can also explain the observed phenomenon. In order to differentiate the minijet picture from the other scenarios, one must find a way to confirm the influence of minijets on particle production.

Although minijets are anticipated from the prediction of perturbative QCD (PQCD) and many efforts have been made to investigate their phenomenological consequences, there exists little direct experimental evidence of their presence in hadronic collisions. The only experimental attempt to find minijets is by UA1 experiments[17] at CERN where hadronic clusters with transverse energy $E_T \gtrsim 5$ GeV have been identified as minijets and the cross sections are found to be consistent with the PQCD prediction. However, this kind of cluster-finding method even for intermediate $E_T$ values is complicated by the background of random fluctuations[18]. For smaller $E_T$, minijet clusters are overwhelmed by the background fluctuation. Therefore, because of their small transverse momenta, minijets with relatively small $p_T \gtrsim 2$ GeV/c can never be resolved as distinct jets from the large soft background in minimum biased events. To avoid the experimental difficulties of reconstructing jets with the huge background in high energy $AA$ collisions, it has been suggested[19] that single inclusive spectra of produced particles can be used to study the properties of jet production. Similarly, I have proposed in a previous paper[20] that two particle correlation functions is useful to study the content of minijets in minimum biased events of hadronic interactions. In this paper, I report the detailed study of the $p_T$ and energy dependence of the two-particle correlation in the azimuthal angle $\phi$ and the influence of finite rapidity cuts.

Since particles from jet fragmentation are correlated in both directions of the minijets, they have been found[21] to dominate both the short and forward-backward two-particle correlations in pseudorapidity. For two-particle correlation in the azimuthal angle $\phi$, contribution from back-to-back minijets should be strongly peaked at both forward ($\Delta\phi = 0$) and backward ($\Delta\phi = \pi$) directions. If we calculate the same correlation, but for some selected particles whose transverse momenta are larger than a certain $p_T$ cut, the two peaks should be more prominent because these particles are more likely to come from minijets. The energy and $p_T$ dependence of such correlations of charged particles must depend on the relative fraction of minijet events and minijet multiplicity, thus shedding some light on minijet content in these events. On the contrary, particles from soft production of fused strings or an expanding quark gluon plasma are isotropical in the transverse plane and would only have some nominal correlation in the backward direction due to momentum conservation.

The experimental study of the two-particle correlations is important to clarify the controversial issue whether the observed flavor dependence of the correlation between the multiplicity and average transverse momentum at Fermilab Tevatron energy[13] is due to minijet production[12], or string interaction[16], or the formation of a quark gluon plasma[14]. On the other hand, it can also provide constraints on theoretical models as how to combine PQCD hard scatterings with nonperturbative soft interactions, how to parametrize the soft processes, and what is the appropriate fragmentation scheme of multiple minijets. In the light of the study on intermittency in particle production[22], it also helps us to have a better understanding of the intermittent fluctuations in hadronic collisions at collider energies as particles from jet fragmentation indeed have been shown to have strong intermittent behavior[23].

This study is based on the Monte Carlo model, HIJING[24], which combines a simple two-string phenomenology for low $p_T$ processes together with PQCD for high $p_T$ processes. Since HIJING is a model with a very specific scheme of minijet fragmentation, the results we get are only a qualitative estimate of the effects of minijets on the two-particle correlation functions. Though the correlations in pseudorapidity $\eta$ and azimuthal angle $\phi$ are both important, we limit our discussions only to the latter case. The remaining of this paper is organized as follows. In Sec. II we briefly review the HIJING model as discussed in detailed in Ref.[24]. In Sec. III we study the $p_T$ dependence of the two-particle correlations in $\phi$ and their resemblance to large $p_T$ jet profiles. Special emphasis is given to the effects of minijets on the back-to-back differences of the correlation, $c(0,0) - c(0,\pi)$. In Sec. IV, the energy dependence of the background of the correlation and its implication on multiple minijet production are discussed. The effect of finite rapidity cut is also discussed. Finally, Sec. V concludes with a summary and remarks.

## II. THE HIJING MODEL

The HIJING model[24] has been developed mainly for multiple jet and particle production in $pp$, $pA$, and $AA$ collisions at energies $\sqrt{s} \gtrsim 5$ AGeV. The formulation of HIJING was guided by the successful implementation of PQCD in the PYTHIA model[25] and the need to incorporate a consistent model of soft processes. It thus provides a link between the dominant nonperturbative fragmentation physics at intermediate energies and the perturbative physics at higher collider energies. A detailed description of the HIJING model can be found in Ref.[24]. It includes multiple jet production with initial and final state radiation along the lines of the PYTHIA model[25] and soft beam jets are modeled by quark-diquark strings with gluon kinks along the lines of the DPM[26] and FRITIOF[27] models. For nucleus induced reactions, nuclear shadowing of parton structure functions and final state interaction of produced jets are also considered[19].

In the eikonal formalism, the inelastic cross section of $pp$ or $p\bar{p}$ collisions is given by

$$\sigma_{in} = \int d^2b [1 - e^{-(\sigma_{soft}+\sigma_{jet})T_N(b,s)}], \tag{1}$$

where $T_N(b,s)$ is the partonic overlap function between two nucleons at impact parameter $b$, $\sigma_{jet}(p_0, s)$ is the total inclusive jet cross section calculated from PQCD with $p_T > p_0$, and $\sigma_{soft}(s)$ is the corresponding phenomenological inclusive cross section of soft interactions. The cross sections for no and $j \geq 1$ number of hard or semihard parton scatterings are,

$$\sigma_0 = \int d^2b [1 - e^{-\sigma_{soft}T_N(b,s)}]e^{-\sigma_{jet}T_N(b,s)}, \tag{2}$$

$$\sigma_j = \int d^2b \frac{[\sigma_{jet}T_N(b,s)]^j}{j!}e^{-\sigma_{jet}T_N(b,s)}, \tag{3}$$

with their sum giving rise to the total inelastic cross section, $\sigma_{in}$. Following the above formulas, the cross section with at least one hard parton scattering, defined as $\sigma_{hard}$, is then

$$\sigma_{hard} = \int d^2b [1 - e^{-\sigma_{jet}T_N(b,s)}] e^{-\sigma_{soft}T_N(b,s)}, \qquad (4)$$

and the average number of parton scatterings per inelastic event is

$$\langle n_{jet} \rangle = \sigma_{jet}(p_0, s)/\sigma_{in}. \qquad (5)$$

Shown in Fig. 1 are the energy dependence of $\langle n_{jet} \rangle$ and the fractional cross section of hard processes, $\sigma_{hard}/\sigma_{in}$ for $p_0 = 2$ GeV/c and $\sigma_{soft} = 57$ mb. We see that as energy increases inelastic events contain more and more hard processes and finally are dominated by minijet production. The probabilities for multiple minijet production also become prominent at energies above 1 TeV as $\langle n_{jet} \rangle \gtrsim 1$.

Once again, we emphasize here that $p_0$ and $\sigma_{soft}$ in HIJING model are phenomenological and model dependent parameters separating PQCD at high $p_T$ from the nonperturbative low $p_T$ regime. Our guideline for choosing the value of $p_0$ is that it should be much larger than $\Lambda_{QCD} \sim 200$ MeV/c in order to apply PQCD but low enough to permit the simplest possible model for the nonperturbative dynamics. Choosing $p_0 = 2$ GeV/c allows us to use a *constant* $\sigma_{soft} = 57$ mb to reproduce the observed energy dependence of the total, elastic and inelastic cross sections in $pp$ and $p\bar{p}$ collisions[10, 21]. With these parameters, it has been shown[21] that a simple two-string model suffices to account for the soft dynamics. A complete test of the HIJING model has been reported in Ref. [21]. The model has been successful to provide a consistent explanation of not only the energy dependence of pseudorapidity distributions, moderate $p_T$ inclusive spectra, and the violation of Koba-Nielsen-Olesen (KNO) scaling, but also the flavor and multiplicity dependence of the average transverse momentum[12] in $pp$ and $p\bar{p}$ collisions in a wide energy range $\sqrt{s} = 50$–1800 GeV. In addition, it is also consistent with the available data on $pA$ and $AA$ collisions at moderate energies $\sqrt{s} \lesssim 20$ AGeV[24].

### III. TWO-PARTICLE CORRELATION

Experimental study of particles from the fragmentation of high $p_T$ jets has shown that particles with larger average transverse momenta are much concentrated in both directions of the back-to-back jets. The widths of these jet profiles are about 1 in both pseudorapidity $\eta$ and azimuthal angle $\phi$ as has been determined from the calorimetric study of high $p_T$ jets. It is apparent that similar trends should also hold for particles from minijets and they must influence the two-particle correlation functions. It has already been shown in Ref. [21] that minijets are the dominant mechanism underlying both the enhanced short range and forward-backward correlations in $\eta$. In this paper I demonstrate that the energy and $p_T$ dependence of two-particle correlations in the azimuthal angle $\phi$ can be used to study minijets in minimum biased events. One can expect that with higher $p_T^{\rm cut}$, the correlation must look similar to the high $p_T$ jet profiles. Moreover, the energy dependence of the correlation for fixed $p_T^{\rm cut}$ should be related to the increase of the fraction of minijet events and the average number of minijets as shown in Fig. 1.

The normalized two-particle correlation functions are defined as,

$$c(\phi_1, \phi_2) \equiv \frac{\rho(\phi_1, \phi_2)}{\rho(\phi_1)\rho(\phi_2)} - 1, \tag{6}$$

where $\rho(\phi)$ is the averaged particle density in $\phi$ and $\rho(\phi_1, \phi_2)$ is the two-particle density which is proportional to the probability of joint particle production at $\phi_1$ and $\phi_2$. Both $\rho(\phi)$ and $\rho(\phi_1, \phi_2)$ are integrated over the whole rapidity range. The limitation of a restricted rapidity range will be discussed in the next section. In actual calculations, one direction in the transverse plane is randomly selected as $\phi_1 = 0$ for all the events and then both the averaged single inclusive density $\rho(\Delta\phi)$ and the two-particle joint density $\rho(0, \Delta\phi)$ are calculated. Because the collisions are symmetrical in the transverse plane perpendicular to the beam direction, $\rho(\Delta\phi)$ should be a constant equal to the total averaged multiplicity of the selected particles divided by $2\pi$. Similarly, $\rho(0, \Delta\phi)$ is also independent of the direction which is chosen as $\phi_1 = 0$. In fact, one can even choose the direction of each particle as $\phi_1 = 0$ to calculate $c(0, \Delta\phi)$ and then average it over all particles in each event. In this case one can get better statistics for a limited number of events. Since I have the luxury to produce many events through the Monte Carlo simulation, I instead try to average over as many events as we can to achieve the best statistics possible over global quantities such as the number of jets per event with relative large $p_T$.

Shown in Figs. 2–5 are calculated results on two-particle correlation functions in $pp$ and $p\bar{p}$ collisions at different energies. To study the $p_T$ dependence, the correlations are obtained for selected particles with different transverse momentum cut $p_T > p_T^{\rm cut}=0$ (long-dash-short-dashed histograms), 0.5 (dotted histograms), 1.0 (solid histograms), 1.5 (short-dashed histograms), and 2.0 (dot-dashed histograms). At low energy $\sqrt{s} = 50$ GeV (Fig. 2) where minijet production is not very important as seen from Fig. 1, there is very little correlation between two particles with $\Delta\phi \lesssim 2\pi/3$ except some small increase at $\Delta\phi \sim 0$ for the case of $p_T > 1.0$ GeV/c. The correlation at $\Delta\phi \sim \pi$ is totally due to the momentum conservation which increases with larger $p_T$ cuts. As was pointed out in Ref. [21], minijet contribution to particle production at the highest CERN ISR energies $\sqrt{s} \approx 50$ GeV is just beginning to emerge. This is again illustrated here by the slight increase of the correlation at $\Delta\phi \sim 0$ for particles with $p_T > 1.0$ GeV/c.

As we can see from Fig. 1, minijet production increases very fast with energy and becomes a significant part of the hadronic interactions above $\sqrt{s} \gtrsim 100$ GeV. Therefore, the two-particle correlation functions in Fig. 3 for $\sqrt{s} = 200$ GeV are very different from those at $\sqrt{s} = 50$ GeV in Fig. 2. We observe that there is apparently strong correlation between two particles with $p_T > 0.5$ GeV/c at both $\Delta\phi \sim 0$ and $\pi$ forming a valley at $\Delta\phi \sim \pi/3$. This feature is very similar to the high $p_T$ jet profiles as functions of $\phi$ relative to the triggered jet axis[28]. However, due to the relative large fraction of particles from soft interactions where momentum conservation dictates the two-particle correlation at $\Delta\phi \sim \pi$, the resultant $c(0,\Delta\phi)$ at $\Delta\phi \sim \pi$ is still larger than at $\Delta\phi \sim 0$. Whereas in high $p_T$ jet profiles, the peak of particle flow toward the triggered jet ($\Delta\phi = 0$) is always narrower but the height is about the same as in the backward direction($\Delta\phi = \pi$). The dominance of particles from soft interactions at this energy is clearly demonstrated in the case of $p_T^{\rm cut} = 0$ where $c(0,\Delta\phi)$ monotonically increases from 0.2 at $\Delta\phi = 0$ to 0.3 at $\Delta\phi = \pi$.

As we increase the energy to 1.8 and 6 TeV (Figs. 4, 5), the correlation functions with no $p_T$ cut become flatter over the whole angular range due to the increasing fraction of particles from minijets relative to the soft processes. The correlation functions with higher $p_T$ cuts look more and more similar to the high $p_T$ jet profiles with $c(0,0) \approx c(0,\pi)$ and narrower peaks at $\Delta\phi \sim 0$. We know from Fig. 2 that particles from soft production have some strong correlation only at $\Delta\phi \sim \pi$ due to momentum conservation while particles from minijets, on the other hand, have enhanced correlation both at $\Delta\phi \sim 0$ and $\pi$. The actual shape of the correlation function then depends on the relative fractions of particles from minijets and soft processes. Looking through Figs. 2–5, we can clearly see that the height of the correlation functions at $\Delta\phi \sim \pi$ gradually becomes equal to that at $\Delta\phi \sim 0$ as shown in Fig. 6 by the increase of the difference $c(0,0) - c(0,\pi)$ with energy. It approaches zero for any fixed $p_T^{\rm cut}$ at higher energies. This energy dependence of $c(0,0) - c(0,\pi)$ is a direct consequence of the increase of minijet production with energy in $pp$ and $p\bar{p}$ collisions. Thus, it has been demonstrated that the study of the energy and $p_T$ dependence of two-particle correlation function $c(0,\Delta\phi)$ can provide us with information about the minijet content in minimum biased events.

## IV. EFFECTS OF MULTIPLE MINIJETS AND FINITE RAPIDITY CUT

When the average number of produced minijets is larger than one (Fig. 1) at high colliding energies, multiple minijet production becomes important[10, 24]. Since these minijets are independently produced, they should increase the background to the two-particle correlation. The very extreme cases are central events of high energy heavy ion collisions where hundreds of minijets are produced. The two-particle correlation will be completely flat when the valleys we see in Figs. 2–5 for $pp$ and $p\bar{p}$ collisions are filled up by the increased background. In hadronic collisions this rarely happens and the increase of the background can be considered as an implication of multiple minijet production.

Plotted in Fig. 7 is the difference $c(0,0) - c(0,\pi/3)$ between the correlation at $\Delta\phi = 0$ and the background in the valley $\Delta\phi = \pi/3$ for each fixed $p_T^{\text{cut}}$ as a function of $\sqrt{s}$. It grows first with energy due to the increase of minijet events and the relative small number of minijets per event. However, as the probabilities of multiple minijet production become increasingly important, the difference then decreases with energy. The larger the $p_T^{\text{cut}}$, the sooner the decrease begins.

Unlike high $p_T$ back-to-back jets which are both kinematically bounded to the central rapidity region, a pair of minijets can be easily produced with large rapidity gap between them. When we trigger one minijet in a limited rapidity window, the other one which is produced in the same parton scattering often falls outside the rapidity window. Therefore, if we calculate two-particle correlation for particles in a limited rapidity range, minijet contribution to the backward correlation ($\Delta\phi = \pi$) will mostly drop out while the contribution to forward correlation ($\Delta\phi = 0$) still remains. Indeed, as shown in Fig. 8, the forward correlation at $\Delta\phi = 0$ for particles in $|\eta| < 1$ is very strong, but the backward correlation at $\Delta\phi = \pi$ is drastically reduced as compared to the correlation pattern in the full rapidity range in Fig. 2-5. Furthermore, due to strong short range two-particle correlation in rapidity[21], the forward correlation at $\Delta\phi = 0$ is enhanced by restricting particles to $|\eta| < 1$. At $\sqrt{s} = 1.8$ TeV, I find that the enhancement of backward correlation at $\Delta\phi = \pi$ due to minijets becomes important only when the rapidity window is $|\eta| \gtrsim 2$.

## V. SUMMARY AND REMARKS

It is demonstrated in detail in this paper that two-particle correlation functions in the azimuthal angle $\phi$ are useful for the study of minijets and their contribution to particle production in the minimum biased events of hadronic interactions. Because particles from jet fragmentation are more concentrated along the jet axes, they result in enhanced two-particle correlation in both the triggered ($\Delta\phi = 0$) and the backward ($\Delta\phi = \pi$) direction. On the other hand, particles from isotropic soft production only have some nominal correlation in the backward direction due to momentum conservation. Therefore, the energy dependence of the relative height $c(0,0) - c(0,\pi)$ of the correlation in the two opposite directions is directly related to the minijet content in the particle production and increases with the colliding energy. Furthermore, multiple minijet production becomes important at high energies and they increase the background relative to the enhanced correlation at $\Delta\phi = 0$ and $\pi$.

It is important to keep in mind that the results we obtained via HIJING are only intended to show qualitatively the effects of multiple minijets on the energy and $p_T$ dependence of two-particle correlation in $\phi$. The quantitative magnitudes of the correlations must depend on the specific scheme of minijet fragmentation and the color flow of multiple minijets[24] in HIJING. It is important to study in the future the sensativity of two-particle correlation functions to the schemes of minijet fragmentation and the color flow structure of multiple minijets. The final schemes which survive the tests on single inclusive spectra have also to be constrained by the experimental measurements of the energy and $p_T$ dependence of two-particle correlation functions.

I have emphasized the implications of the energy and $p_T$ dependence of the two-particle correlation functions on minijet production. For experiments at a fixed energy, one can also investigate the contribution of minijets to the two-particle correlation function by comparing the direct measurements to the calculation from the same data sample but with randomized azimuthal angle $\phi$ for each particle. Due to this randomization of $\phi$, the correlation calculated from the manipulated data should be much flatter than the real one for any given $p_T^{\text{cut}}$. It is also important and interesting to study the $p_T$ dependence of two-particle correlation functions in pseudorapidity $\eta$ and the correlation relative to a triggered particle which has the highest $p_T$ in each event[20].

As discussed in Sec. IV, the two-particle correlation in $\phi$ in ultrarelativistic heavy ion collisions should be much flatter than in $pp$ and $p\bar{p}$ collisions for any fixed $p_T^{\text{cut}}$ due to the huge number of minijet production. However, if we increase $p_T^{\text{cut}}$ to a much larger value (e.g. 4–6 GeV/c), the background to the correlation becomes smaller due to small cross sections of large $p_T$ jets. Since the single inclusive spectrum at large and moderate $p_T$ has been shown to be sensitive to jet quenching[19], two-particle correlation is apparently more relevant to the study of jet quenching and the implication on energy loss mechanism it conveys.

## ACKNOWLEDGMENTS

I would like to thank A. Goshaw, M. Gyulassy, B. Müller, and S. Oh for their helpful comments and encouraging discussions. This work was supported by the Director, Office of Energy Research, Division of Nuclear Physics of the Office of High Energy and Nuclear Physics of the U.S. Department of Energy under Contract No. DE-AC03-76SF00098 and DE-FG05-90ER40592.

FIGURES

FIG. 1. Fractional cross section $\sigma_{hard}/\sigma_{in}$ for at least one hard scattering and average number of minijet production $\langle n_{jet}\rangle$ per event as functions of $\sqrt{s}$.

FIG. 2. The correlation functions $c(0,\Delta\phi)$ between two particles with fixed $p_T$ cutoff vs $\Delta\phi$ in $pp$ collisions at $\sqrt{s} = 50$ GeV.

FIG. 3. The same as Fig. 2, except for $p\bar{p}$ collisions at $\sqrt{s} = 200$ GeV.

FIG. 4. The same as Fig. 2, except for $p\bar{p}$ collisions at $\sqrt{s} = 1.8$ TeV.

FIG. 5. The same as Fig. 2, except for $\sqrt{s} = 6$ TeV.

FIG. 6. The energy dependence of the difference $c(0,0) - c(0,\pi)$ between the correlation at $\Delta\phi = 0$ and $\pi$ for each fixed $p_T$ cutoff.

FIG. 7. The energy dependence of the difference $c(0,0) - c(0,\pi/3)$ between the correlation at $\Delta\phi = 0$ and the background at $\pi/3$ for each fixed $p_T$ cutoff.

FIG. 8. The same as Fig. 4, except for charged particles in a limited rapidity range of $|\eta| < 1$.

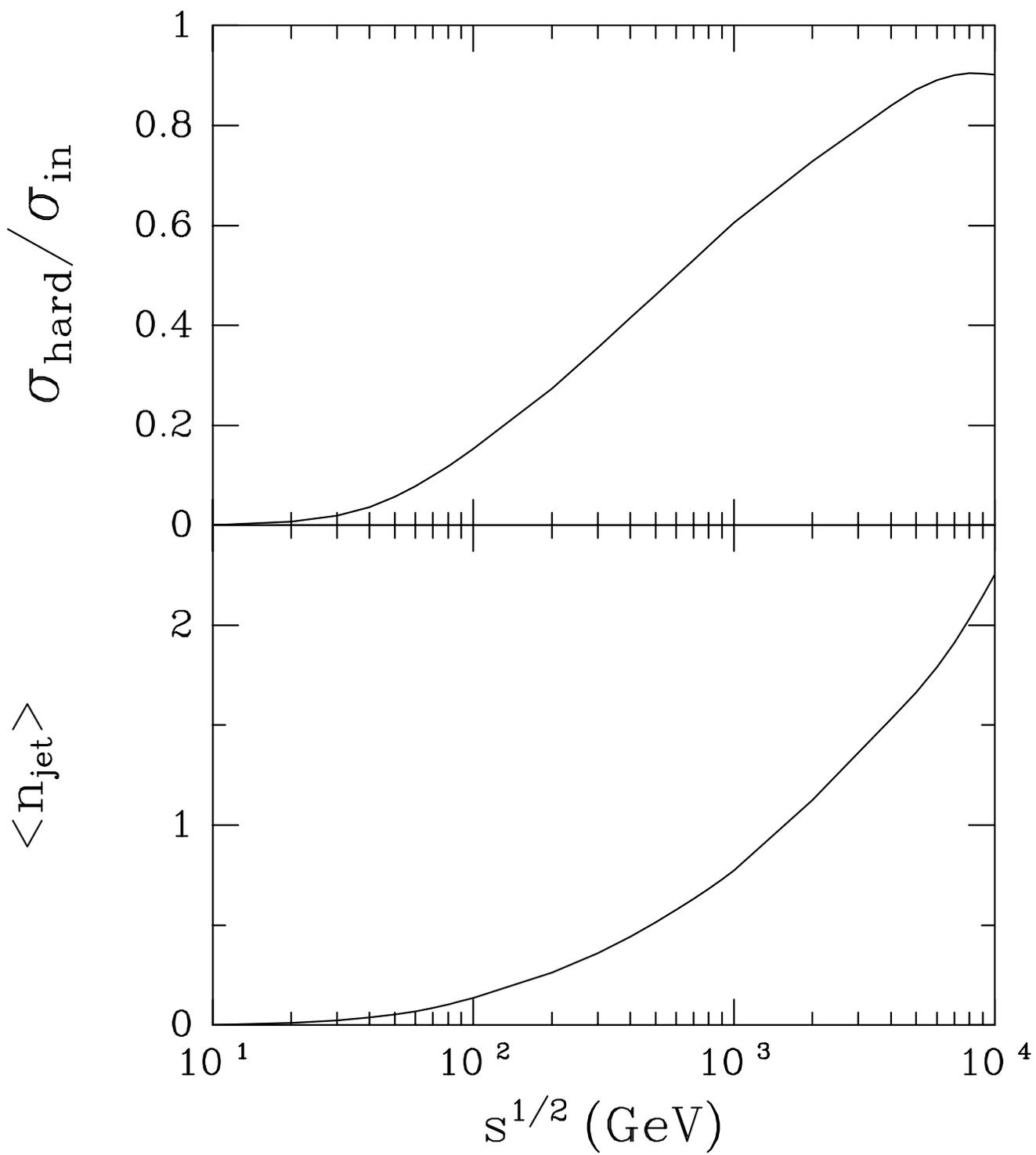

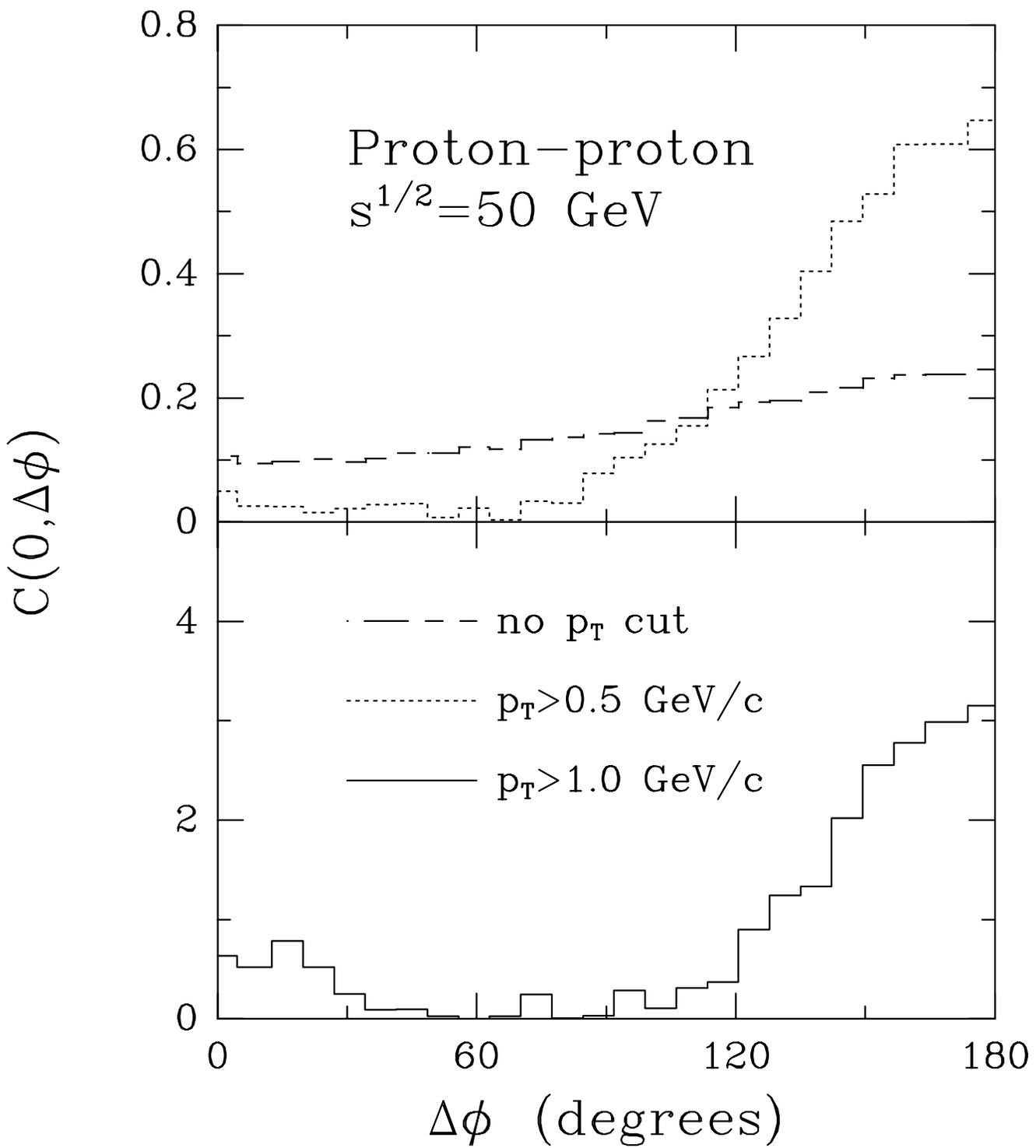

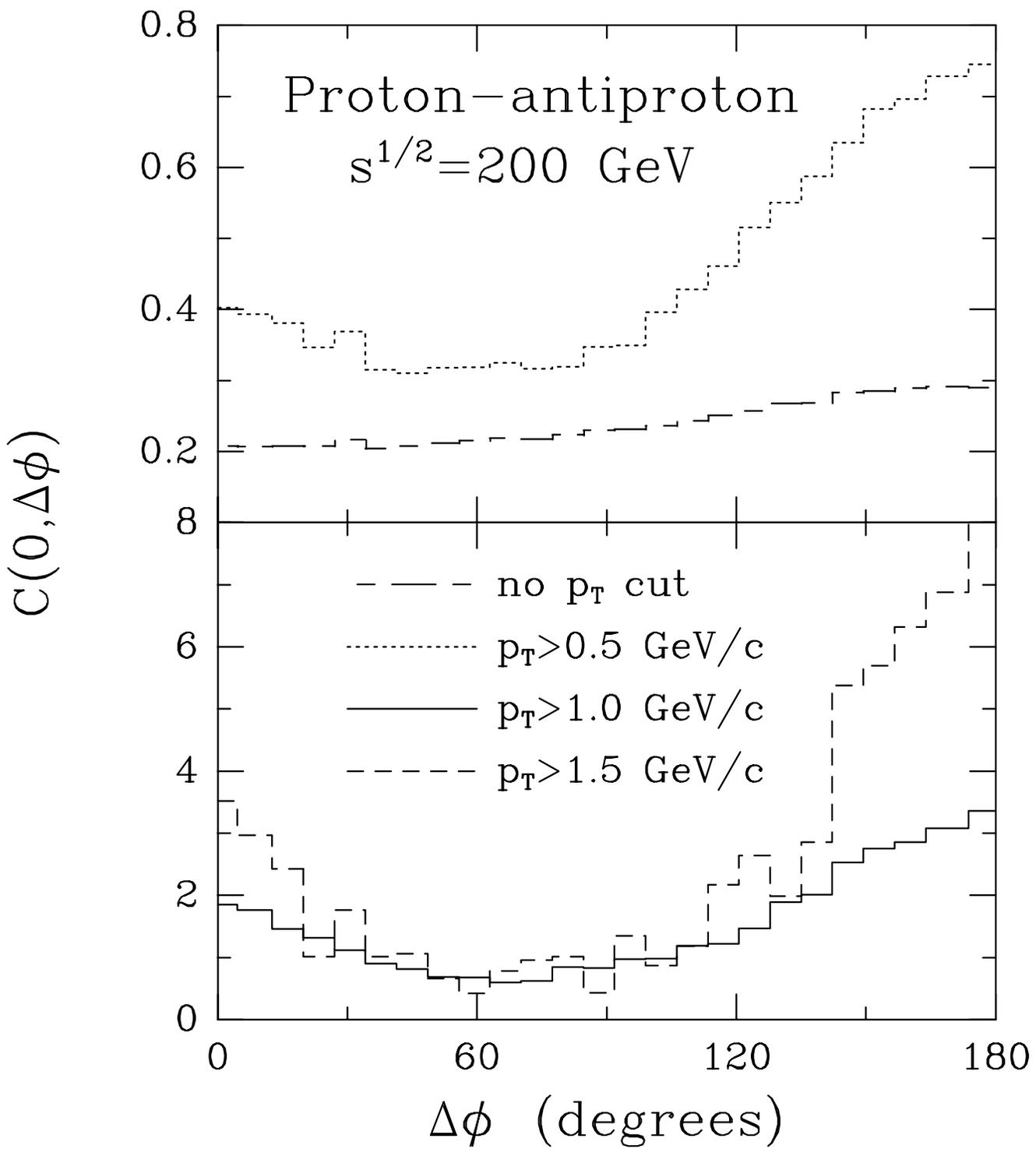

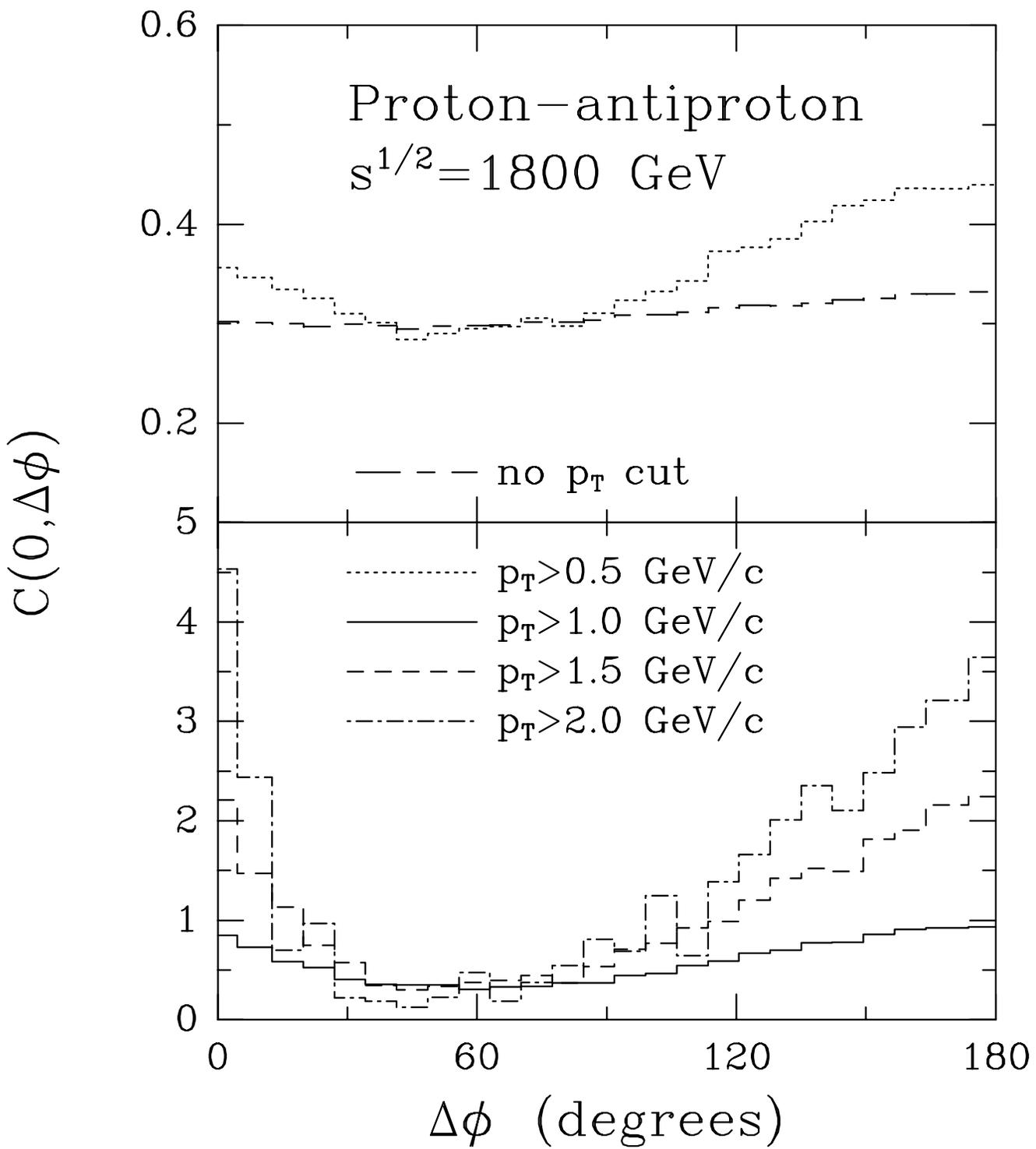

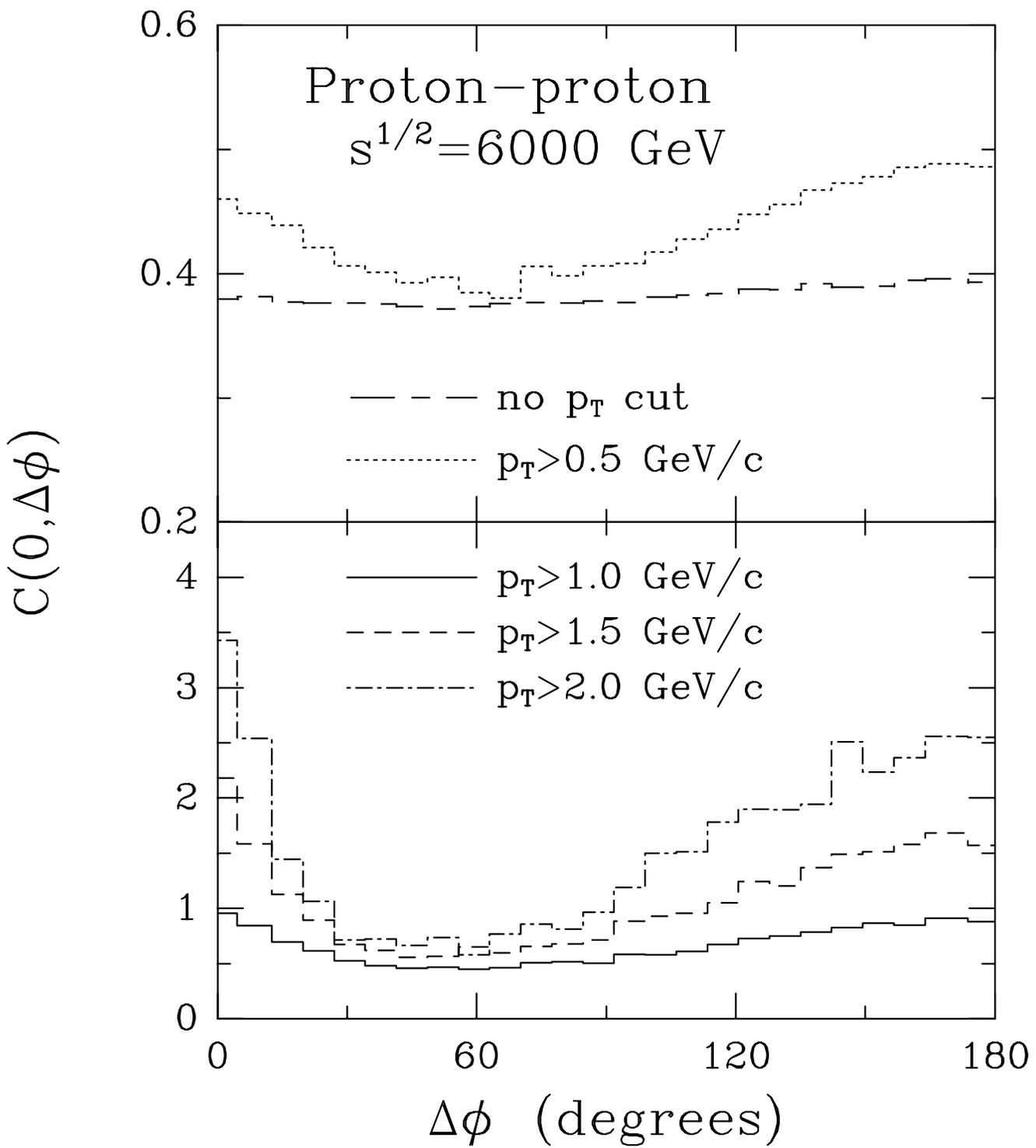

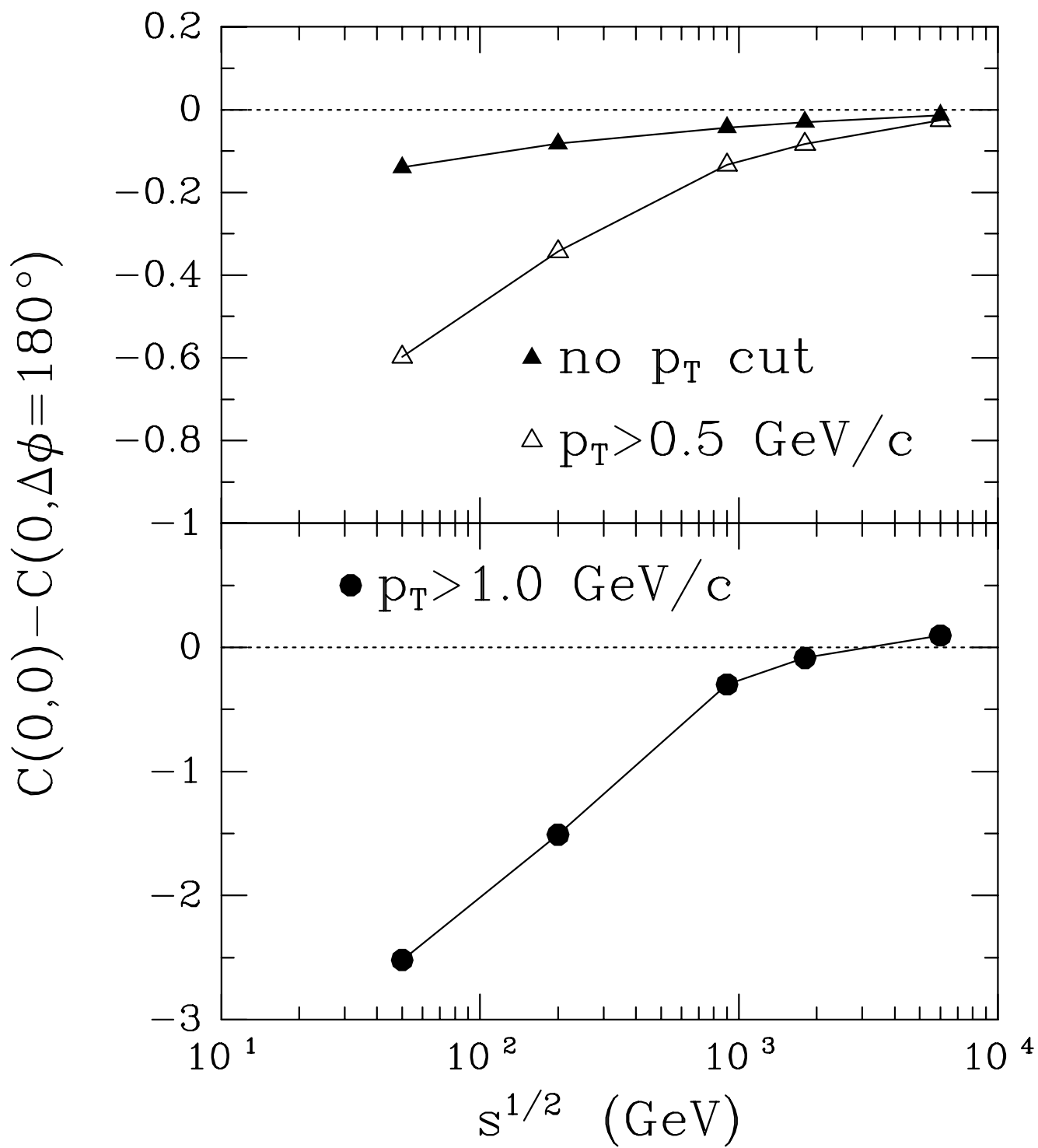

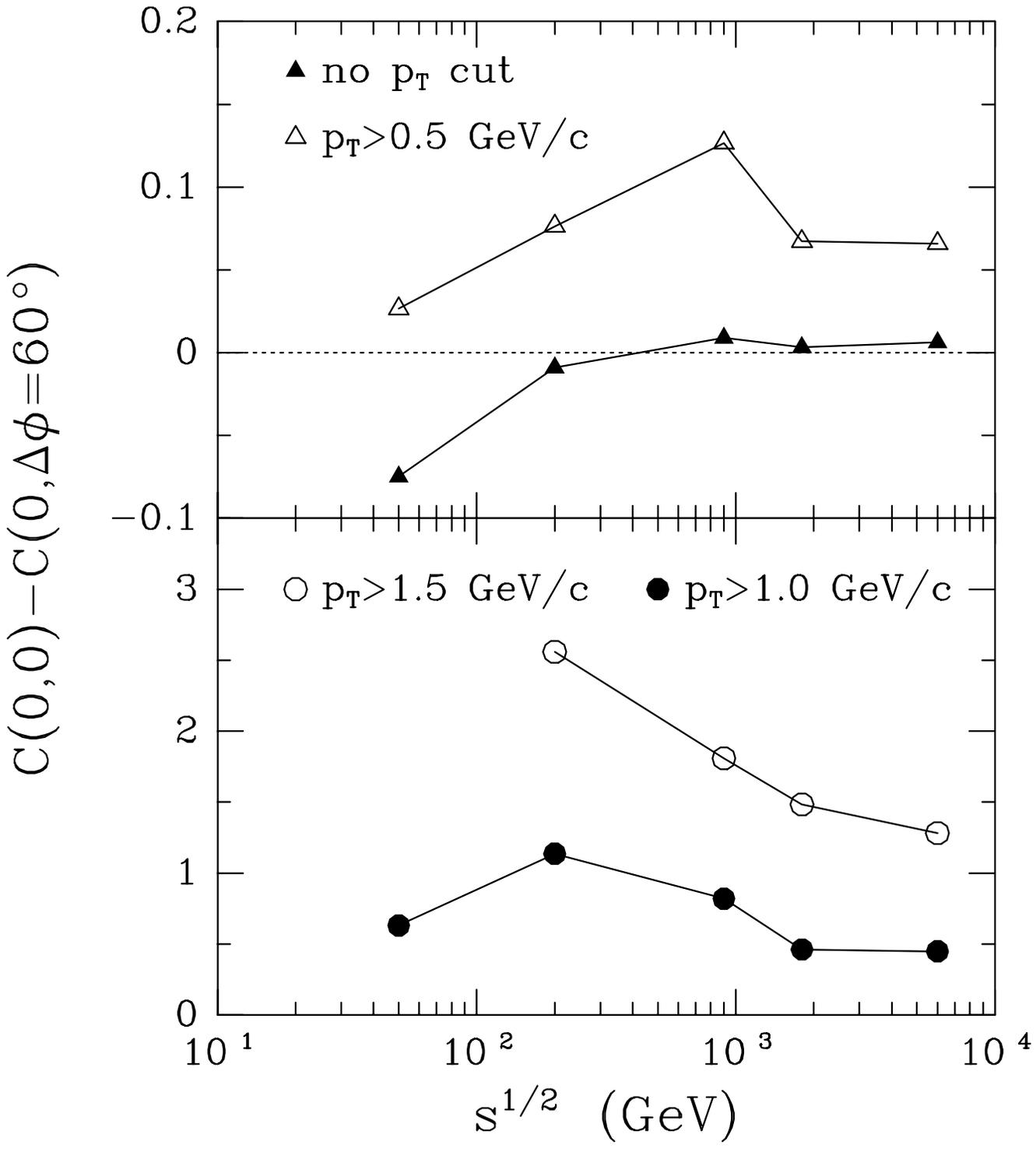

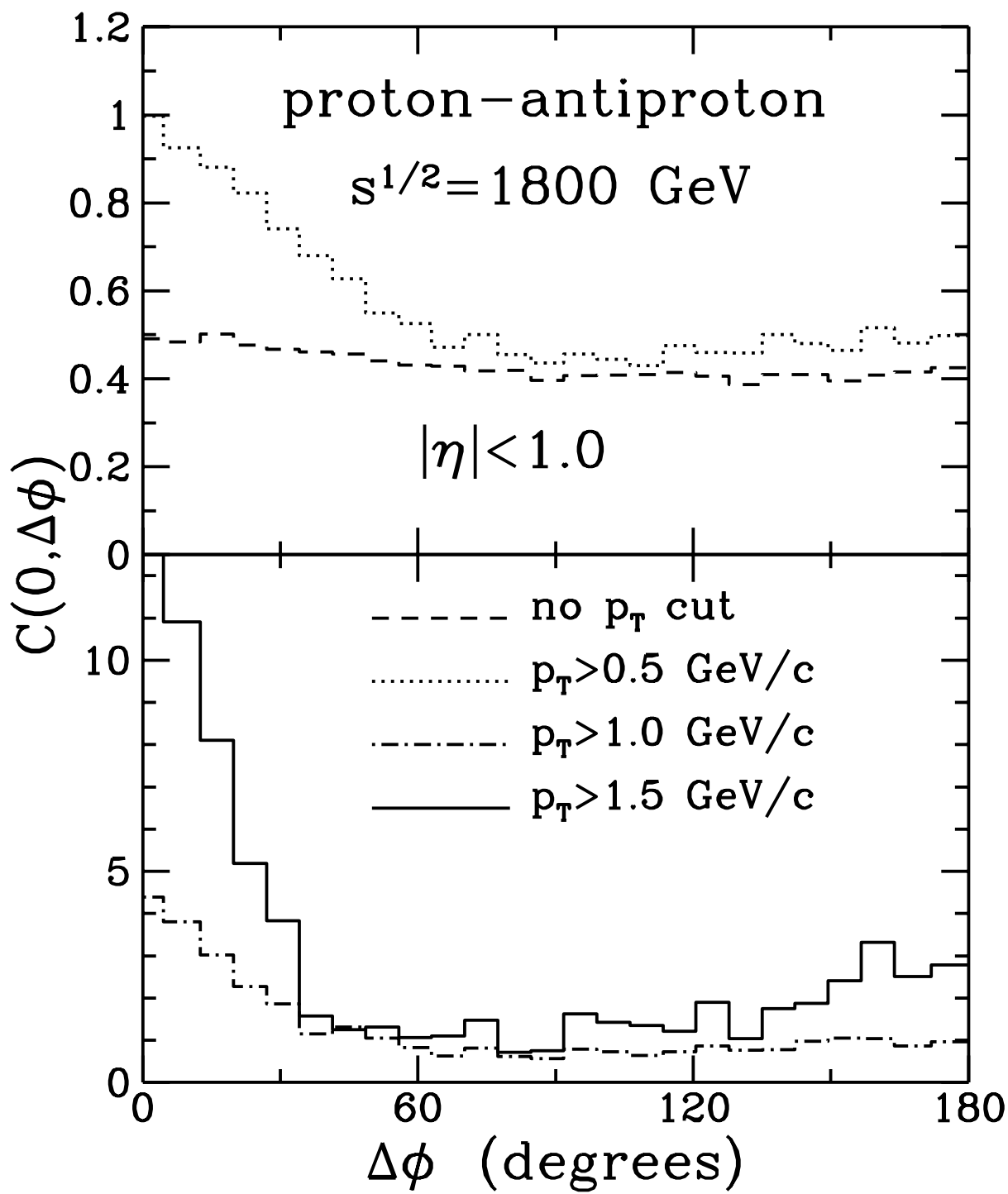